# BUDDEN'S PARADOX RESOLVED

J. A. Grzesik
Allwave Corporation
3860 Del Amo Boulevard
Suite 404
Torrance, CA 90503

(818) 749-3602
jan.grzesik@hotmail.com

September 20, 2018

**Abstract**

The Budden energy nonconservation paradox is dispelled herein by recognizing that pole approach to the spatial origin from below in the complex plane can be resolved into a real principal value minus $i\pi$ times a Dirac delta, the imaginary coefficient whereof supplies just the right amount of localized dissipation to equilibrate the energy budget precisely, regardless of whether wave incidence be up or down. Only the reflectionless downward wave incidence remains as a counterintuitive challenge to physical anticipation, but at least a challenge liberated from its former onus of energy nonconservation.

*Key Words*–Budden's energy nonconservation paradox, wave transit across a refractive index singularity, Cauchy principal value/Dirac delta canonical singularity recipe, Whittaker function values both remote and close in, Ohmic/Joule energy dissipation

# 1　Introduction

One celebrated legacy of Budden's well known books [1-2] on ionospheric radio wave propagation[1] has been a paradox asserting that a certain case of resonance transit, be it from above or below, fails to equilibrate its energy budget, to the effect that the sum of squared absolute transmission $T_{\pm}$ and reflection $R_{\pm}$ coefficients falls short of unity, *viz.*,

$$|R_{\pm}|^2 + |T_{\pm}|^2 < 1\,, \tag{1}$$

and all of this in the absence of any overt mechanism of dissipation.[2] This unsettling result has been quoted *verbatim*, and rather uncritically, it would surely seem, in [3], while strenuous efforts, including Budden's own discussion, to argue that deficit away pop up sporadically in the literature, in [4-7], and doubtless elsewhere.

In this short note we propose to reveal that defect (1) is fully remedied as soon as we recognize the presence of a Dirac delta conductivity *pulse* highly localized around the refractive index singularity at coördinate origin $z=0$.[3] Indeed, since that singularity has the aspect of $(z+i\gamma)^{-1}$, with $\gamma \downarrow 0+$ measuring a vestigial dissipation due to electron/ion background collisions, it follows that

$$\frac{1}{z} \doteq \lim_{\gamma\downarrow 0+}\frac{1}{z+i\gamma} = \frac{P}{z} - i\pi\delta(z) \tag{2}$$

whereby we unmask a dissipative term $-i\pi\delta(z)$ which had, so to speak, been hiding in the open all along (symbol $P$ denotes the Cauchy principal value). And then, in a standard interplay of the Ampère and Faraday equations, we can interpret the imaginary part of (2) as a conductivity *pulse* $\sigma(z)$.[4] That conductivity, as will shortly become apparent, is of just the right magnitude to account for the apparent energy defect in both upward and downward wave passage, and thus, hopefully at long last, to downgrade the notoriety of Budden's paradox.

It must be kept in mind that Budden's propagation model confines itself to a very simple physical microcosm. Complex phenomena such as mode conversion, while they and their attendant mathematics exercise legitimate rôles in nature, simply lie outside its purview. It must in particular admit a purely internal validation of its energetic self-consistency, with energy up/down budgets balanced on their own merits, without futile, frenzied attempts to seek exterior recourse in mode conversion, all of which should be deemed as little more than counsels of desperation, earnest, honest, and sophisticated though their goals and methods may be.

---

[1]While Budden [1] is admittedly archival, Budden [2] is of a considerably later vintage, and should prove thus to be more readily accessible. It is the latter which is the target of our detailed equation number citations that follow.

[2]Subscripts $\pm$ correspond to upward/downward transit across the resonance singularity.

[3]We adhere to SI units, so that $z$ is measured in meters and electric field $E_y$ (Eq. (3) onward) in volts per meter.

[4]Additional multipliers are best deferred to Eq. (5) following. We intend to adhere closely to Budden's own notation so as not to encumber still further a radio wave theory already mired in a prolix avalanche of symbols. In particular, in keeping with Budden's convention, electric fields are considered to depend on time $t$ in accordance with $e^{i\omega t}$, $\omega > 0$, neither $t$ nor $e^{i\omega t}$ being explicitly mentioned. Equations from [2], our main literature contact, are signalled with a prefix B for Budden, *viz.*, Eq. (B19.55) and so forth.

One example of such efforts can be found in [6], and in its subsequent refinement [7], both of which seek to cast Budden's energy defect beneath the guise of an Alfvén into ion-Bernstein wave conversion. But this is patently incongruous, since ion-Bernstein waves, while a most valid phenomenon in their own right, are altogether shielded from view by the self-imposed curtains of the Budden model, an energy refuge forbidden by formal fiat. Grafting plasma mode mixing onto a thin Budden substrate cannot do aught but give an impression of straining at the leash.

All of this is not in any way to imply that there is no such phenomenon as plasma mode conversion. On the contrary! It is simply the case that mode conversion must be couched in a much more robust theoretical framework. The level of mathematics which such a task elicits can be traced, for example, from [8].

In the present note we exhibit a very simple, analytically most modest source of dissipation, rooted in nothing more than the Ohmic conductivity due to a vestigial background collisionality, which, in Eqs, (15) and (21) below, *does* balance both up/down energy budgets. Indeed, the analysis is so simple that it should, by all rights, be bold enough to speak for itself.

## 2 Analytic framework

We adopt
$$\frac{d^{2}E_{y}}{dz^{2}} + \left( \lim_{\gamma \downarrow 0+} \frac{k\beta}{z + i\gamma} + \frac{k^{2}\beta^{2}}{\eta^{2}} \right) E_{y} = 0 \tag{3}$$

(Eq. (B19.55) slightly rewritten[5]) as our governing equation, with wavenumber $k = \omega/c$ gotten as the standard ratio of angular frequency $\omega$ to the speed of light $c$. Dimensionless parameters $\beta$ and $\eta$, both real and positive, provide the freedom to fix at $n_{\infty} = \beta/\eta$ the asymptotic limit of the refractive index $n(z)$ (and thus also the asymptotic wavelength $\lambda_{\infty} = 2\pi\eta/k\beta$), to set at $\beta/k$ the strength of the $z = 0$ resonance disturbance of $n^{2}(z)$, and finally to displace in an amount $\Delta z = -\eta^{2}/k\beta$ the refractive index null below the resonance. In this regard one may or may not judge as quaint Budden's choice of notation, but there it is. Limit enforcement in the direction of null collisionality, in accordance with recipe (2),[6] recasts (3) as

$$\frac{d^{2}E_{y}}{dz^{2}} + \left( k\beta\frac{P}{z} - i\pi k\beta\delta(z) + \frac{k^{2}\beta^{2}}{\eta^{2}} \right) E_{y} = 0 \tag{4}$$

where, as already indicated, $P$ stands for the Cauchy principal value and $\delta$ for the Dirac delta. Now, an imaginary term such as $-i\pi k\beta\delta(z)$ in (4) can emerge from the underlying Ampère and Faraday equations only if we acknowledge the existence of an effective conductivity

$$\sigma(z) = \frac{\pi k\beta}{\omega\mu}\delta(z), \tag{5}$$

---

[5]We have bypassed Försterling's $\mathcal{F}$ in the original version of (B19.55) (*q.v.* an appropriate bibliographic Försterling trace within [2]) in favor of the more physically relevant electric field component $E_y$, a step clearly permitted by the first, linear relationship indicated under (B19.69).

[6]Strictly speaking, our $\gamma$, having a dimension of length, is Budden's dimensionless $\gamma$ divided by $k$. As a null limit $\gamma \downarrow 0+$ is being pursued, this technical gloss is without any consequence.

with $\mu$ being the magnetic permeability of the ambient medium, presumably close to $\mu_0 = 4\pi \times 10^{-7}$ H/m. This newfound tool will now unlock the energy deficit puzzle.[7]

To be sure, one may initially recoil from entertaining the existence of a singular conductivity pulse such as (5). But this urge to evade and reject should be tempered by the observation that we already accept, in some sense, the physical existence of Budden's singular refractive index which, in accordance with (2), automatically spawns (5) as the background collisionality recedes to a vanishing point. Of course, away from the resonance, with $|z| > 0$, that same limit is far more benign, leading to up/down nondissipative wave propagation (under the control of Whittaker functions $W$), but with a vestigial, generally unstated damping always hovering in the back of our minds. And in any event, Dirac's delta $\delta$ is nothing other than a convenient shorthand for a limit of otherwise continuous, albeit increasingly sharp analytic entities[8] (and its coefficient $-i\pi$ in (2) arises, equivalently, by retaining the negative of just one half a residue when a left to right contour deforms upward so as to evade a simple pole encroaching from below).

## 3 Up/down propagation scenarios

Equations (3)-(4) entail a self-evident physical asymmetry in the sense that upward wave incidence across resonance singularity at $z = 0$ must first cross at $z_0 = -\eta^2/k\beta$ a refractive index null, whereas the sequence is obviously reversed during downward passage. Budden succeeds in capturing the consequence of such asymmetry by writing solutions of (3) in terms of Whittaker functions $W_{\mp i\eta/2,\pm 1/2}(\pm 2ik\beta z/\eta)$ [**9**], upper/lower signs[9] holding respectively for upward/downward resonance crossing, and then exploiting their dissimilar, $\propto \exp(\mp ik\beta z/\eta)$ asymptotic behaviors following transit.[10] Neither Budden nor we need be bothered to assign any specific inbound asymptotic amplitudes since, in this linear setting, they are all destined to be normalized out. In point of fact, normalization magnitudes are set by an interplay of the asymptotic forms which Whittaker functions acquire, (B19.59)-(B19.61) for wave incidence from below, (B19.66)-(B19.67) for incidence from above. We ourselves utilize these same asymptotic incoming magnitudes when setting Poynting vector asymptotic strengths $S_\pm$ respectively in (11) and (17).

---

[7]How it is that collisions can be identified with pure imaginary additions to dielectric polarization $\boldsymbol{P}$ (not to be confused with the Cauchy principal value from (2)), and thus with dissipative Ohmic currents, can be traced from Eq. (B3.13) and Budden's discussion surrounding it, both fore and aft. In particular, by tracking the algebraic details of that dielectric polarization as found in (B3.14), one duly arrives at the *negative* sign which in (4) is assigned to its Dirac delta term, and which is of the essence in providing a *bona fide* energy sink in (6).

[8]One need only recall that $(z+i\gamma)^{-1} = (z-i\gamma)/(z^2+\gamma^2)$ whereas $\gamma \int_{-\infty}^{\infty} (z^2+\gamma^2)^{-1} dz = \pi$, regardless of how small $\gamma$ becomes.

[9]The sign of the second index, $\pm 1/2$, is discretionary.

[10]Asymptotic phases for $W_{\mp i\eta/2,\pm 1/2}(\pm 2ik\beta z/\eta)$ as catalogued beneath (B19.59), (B19.61), and (B19.66)-(B19.67), include the further terms $\mp i\eta \log(k|z|)/2$ which, following their encounter with a $d/dz$ derivative filter as in (8), are clearly without bearing upon the magnetic field components found in (9) and (16).

### 3.1 Dirac delta dissipation at origin $z = 0$

From (5) there follows[11]

$$
\begin{aligned}
D &= \frac{\pi k \beta}{2\omega\mu} \left| W_{\mp i\eta/2, \pm 1/2}(0) \right|^2 \int_{-a}^{b} \delta(z) dz \\
&= \frac{\pi k \beta}{2\omega\mu} \left| W_{\mp i\eta/2, \pm 1/2}(0) \right|^2
\end{aligned}
\quad (6)
$$

as the time averaged rate of energy dissipation per unit area transverse to propagation direction $z$. And then from [10] and [11] sequentially invoked we get

$$
\begin{aligned}
D &= \frac{\pi k \beta}{2\omega\mu} \left| \Gamma(1 \pm i\eta/2) \right|^{-2} \\
&= \frac{\beta}{2c\eta\mu} \left( e^{\pi\eta/2} - e^{-\pi\eta/2} \right).
\end{aligned}
\quad (7)
$$

On its face, this simple structure does not discriminate as to the $\pm$ directionality of resonance crossing. An obligatory distinction does however rise to the surface in (12) and (18) once the respective $\pm$ normalizers $\propto e^{3\pi\eta/2}$ and $\propto e^{\pi\eta/2}$ have been duly divided out.

### 3.2 Upward resonance crossing

Accompanying an electric component $E_y(z)$ is the single magnetic component

$$
B_x(z) = -\frac{i}{\omega} \frac{dE_y}{dz} \quad (8)
$$

so that, with an upward incident propagation $\propto \exp(-ik\beta z/\eta)$ when $z \to -\infty$,

$$
B_{x,inc}(z) = -\frac{\beta}{\eta c} E_{y,inc}(z). \quad (9)
$$

The time averaged Poynting vector along the direction of increasing $z$ then reads

$$
S_+ = \frac{\beta}{2c\eta\mu} \left| E_{y,inc}(z) \right|^2 \quad (10)
$$

and, on the strength of (B19.61), has the value

$$
S_+ = \frac{\beta}{2c\eta\mu} e^{3\pi\eta/2}. \quad (11)
$$

On dividing $D$ by this latter quantity we get a normalized dissipation

$$
D_+ = e^{-\pi\eta} - e^{-2\pi\eta} \quad (12)
$$

---

[11]Limits $-a$ and $b$ are of course arbitrary, apart from an obvious requirement that $a > 0$ and $b > 0$.

which, in conjunction with

$$|R_+| = 1 - e^{-\pi\eta} \tag{13}$$

(from (B19.62)) and

$$|T_+| = e^{-\pi\eta/2} \tag{14}$$

(from (B19.63)) properly balances the energy budget in the form

$$|R_+|^2 + |T_+|^2 + D_+ = 1 \,. \tag{15}$$

### 3.3 Downward resonance crossing

The kindred calculations are naturally similar, albeit now necessarily anticlimactic. Since the downward incident propagation is proportional to $\exp(ik\beta z/\eta)$, the magnetic field from (9) is obliged to change sign, viz.,

$$B_{x,inc}(z) = \frac{\beta}{\eta c} E_{y,inc}(z) \,, \tag{16}$$

and thus to underwrite a downward energy flow. Poynting vector magnitude $S_-$ remains formally intact as the right-hand side of (10), but with the understanding that such energy flux is a *downflow* in the direction of decreasing $z$. From (B19.66) we encounter the value

$$S_- = \frac{\beta}{2c\eta\mu} e^{\pi\eta/2} \tag{17}$$

whereby $D$ is scaled into

$$D_- = 1 - e^{-\pi\eta} \,. \tag{18}$$

And then, on taking account of the fact that

$$R_- = 0 \tag{19}$$

whereas

$$|T_-| = e^{-\pi\eta/2} \tag{20}$$

(from (B19.72)), it follows once more that

$$|R_-|^2 + |T_-|^2 + D_- = 1 \tag{21}$$

as a reassertion of confidence in energy conservation. That $R_- = 0$ clearly remains as an anomaly for which no physical explanation seems to lie close at hand.

[2] K. G. Budden, **The propagation of radio waves**, *The theory of radio waves of low power in the ionosphere and magnetosphere*, Cambridge University Press, 1988; *Section 19.6. Resonance tunnelling,* pp. 596-602.

[3] Thomas Howard Stix, **Waves in plasmas**, American Institute of Physics Press/Springer-Verlag, New York, 1992; pp. 348-349.

[4] Einar Mjølhus, **Generalized Budden resonance tunnelling, with application to linear conversion nearly parallel to magnetic field**, Journal of Plasma Physics, Volume 38, Issue 1, August 1987, pp. 1-26.

[5] M. C. Williamson, A. J. Lichtenberg, and M. A. Lieberman, **Self-consistent electron cyclotron resonance absorption in a plasma with varying parameters,** Journal of Applied Physics, Volume 72, Number 9, 1 November 1992, pp. 3924-3933.

[6] A. Bers *et al.*, **Mode Conversion of Fast Alfvén Waves to Ion-Bernstein Waves**, MIT Research Laboratory for Electronics Progress Report No. 137, 1994, pp. 229-232.

[7] A. Bers *et al.*, **Mode Conversion of Fast Alfvén Waves to Ion-Bernstein Waves**, MIT Research Laboratory for Electronics Progress Report No. 138, 1995, pp. 248-251.

[8] V. Fuchs, K. Ko, and A. Bers, **Theory of Mode Conversion in Weakly Inhomogenous Plasma**, Physics of Fluids, Volume 24, 1981, pp. 1251-1261.

[9] E. T. Whittaker and G. N. Watson, **A course of modern analysis**, *Fourth edition*, Cambridge University Press, 1962; *Chapter XVI, The confluent hypergeometric function*, pp. 337-354.

[10] Frank W. J. Olver, Daniel W. Lozier, Ronald F. Boisvert, and Charles W. Clark, editors, **NIST handbook of mathematical functions**, National Institute of Standards and Technology, U.S. Department of Commerce, and Cambridge University Press, 2010; *A. B. Olde Daalhuis, Chapter 13, Confluent hypergeometric functions*, p. 335, entry 13.14.17.

[11] Milton Abramowitz and Irene A. Stegun, editors, **Handbook of mathematical functions with formulas, graphs, and mathematical tables**, United States Department of Commerce, National Bureau of Standards, Applied Mathematics Series 55, Tenth Printing, December 1972; *Philip J. Davis, Chapter 6, Gamma function and related functions,* p. 256, entry 6.1.31.